\documentclass[10pt,conference,letterpaper]{IEEEtran}

\IEEEoverridecommandlockouts

\usepackage{times,amsmath,epsfig,amssymb,graphicx,amsfonts,amsthm}
\usepackage{subfigure}
\usepackage{empheq}
\usepackage{psfrag}
\usepackage{fge}
\usepackage{amsmath}
\usepackage{amsthm}
\usepackage{amsfonts}   
\usepackage{amssymb}    
\usepackage{mathrsfs}
\usepackage{setspace}
\usepackage{algorithm}
\usepackage{algcompatible}
\floatname{algorithm}{Routine}

\newtheorem{mytheorem}{Theorem}

\newtheorem{myremark}{Remark}

\newtheorem{myproblem}{Problem}

\newcounter{ale}

\newenvironment{liste}{\begin{itemize}}{\end{itemize}}
\newcommand{\aliste}{\begin{liste} \setcounter{ale}{1}}
\newcommand{\zliste}{\end{liste}}

\newcounter{MYtempeqncnt}

\title{Detection and Isolation of Link Failures under the Agreement Protocol}

\author{M. Amin Rahimian, Victor M. Preciado{\small $~^{*}$}
\thanks{$^{*}$ The authors are with the Department of Electrical and Systems Engineering, University of Pennsylvania, Philadelphia, PA 19104-6228 USA. (email: {\fontsize{8}{8}\selectfont\ttfamily\upshape preciado@seas.upenn.edu}). This work was supported by ONR MURI "Next Generation Network Science".}
}

\begin{document}
\maketitle
\begin{abstract}
In this paper a property of the multi-agent consensus dynamics that relates the failure of links in the network to jump discontinuities in the derivatives of the output responses of the nodes is derived and verified analytically. At the next step, an algorithm for sensor placement is proposed, which would enable the designer to detect and isolate any link failures across the network based on the observed jump discontinuities in the derivatives of the responses of a subset of nodes. These results are explained through elaborative examples.
\end{abstract}

\section{Introduction}
Multi-agent network systems, which consist of a group of interacting dynamic agents, have found promising applications in areas such as motion coordination of robots \cite{mesbahiBook}. Such cooperative dynamics over a network may be strongly affected by the network failures and this has motivated the study of network dynamics following the removal of some links or nodes \cite{aminAutomatica}. By and large, the study of failures is an important topic in network science and it has various practical implications \cite{Kleinberg}. Consensus or agreement protocol has been extensively investigated in the recent literature as a fundamental evolution law for multi-agent networks \cite{RenBeard2008}. The papers \cite{FailureDetectionACC2012} and \cite{FailureDetectionCDC2012} address the issue of detectability for single and multiple link failures in a multi-agent system under the agreement protocol, where it is pointed out that link failures are detectable for a class of directed graphs with rooted out-branchings. Additional conditions in terms of the inter-nodal distances to the observation points are also provided for the detectability of links.  The chief aim of this paper is to provide a method for detection and isolation of single link failures in a network that evolves according to the linear agreement protocol, based on the output responses of a subset of nodes.

The remainder of this paper is organized as follows. Section~\ref{sec:pre} gives some preliminaries on sets and graph theory, and introduces the notation that is used throughout the paper. The main theorem that forms the analytic basis for the proposed detection method is stated and proved in Section~\ref{sec:derivatives}. Next in Section~\ref{sec:sensorpalcement} a set of algorithms are proposed for the effective selection of observation points in the network. These algorithms together with the theorem in Section~\ref{sec:derivatives}, enable the network designer to detect and isolate single link failures based on the observed jump discontinuities in the derivatives of the output responses of a subset of nodes. Illustrative examples and discussions in Section~\ref{sec:examples} elucidate the results and Section~\ref{sec:conc} concludes the paper.

\section{Sets and Digraphs}\label{sec:pre}

Throughout the paper, $\varnothing$ is the empty set, $\mathbb{N}$ denotes the set of all natural numbers,$\mathbb{W} = \mathbb{N} \cup \{0\}$, and $\mathbb{R}$ denotes the set of all real numbers. Also, the set of integers $\{1,2,\ldots,k\}$ is denoted by $\mathbb{N}_k$, and any other set is represented by a curved capital letter. The cardinality of a set $\mathscr{X}$, which is the number of its elements, is denoted by $|\mathscr{X}|$, and $\mathscr{P}(\mathscr{X}) = \{ \mathscr{M}; \mathscr{M} \subset \mathscr{X} \}$ denotes the power-set of $X$, which is the set of all its subsets. The difference of two sets $\mathscr{X}$ and $\mathscr{Y}$ is  denoted by $\mathscr{X} \fgebackslash \mathscr{Y}$ and is defined as $\left\{x;x \in \mathscr{X} \wedge x \notin \mathscr{Y}\right\}$, where $\wedge$ is the logical conjunction. In additional the logical implication and bi-implication are denoted by $\rightarrow$ and $\leftrightarrow$, respectively. Matrices are represented by capital letters, vectors are expressed by boldface lower-case letters, and the superscript $^{T}$ denotes the matrix transpose. Moreover, $I$ denotes the identity matrix with proper dimension, and the determinant of a matrix $D$ is denoted by $\det(D)$, while $\left[D\right]_{ij}$ indicates the element of $D$ which is located at its $i-$th row and $j-$th column. 

A directed graph or \emph{digraph} is defined as an ordered pair of sets $\mathscr{G} := (\mathscr{V},\mathscr{E})$, where $\mathscr{V} = \left\{{\nu}_1,\ldots,{\nu}_n\right\}$ is a set of $n = |\mathscr{V}|$ vertices and $\mathscr{E} \subseteq \mathscr{V} \times \mathscr{V}$ is a set of directed edges. In the graphical representations, each edge $\epsilon := ({\tau},{\nu}) \in \mathscr{E}$ is depicted by a directed arc from vertex ${\tau} \in \mathscr{V}$ to vertex ${\nu} \in \mathscr{V}$. Vertices ${\nu}$ and $\tau$ are referred to as the \emph{head} and \emph{tail} of the edge $\epsilon$, respectively; and if $\tau = \nu$, then $\epsilon$ is dubbed a self-loop.  Given a set of vertices $\mathscr{X} \subset \mathscr{V}$, the set of all edges for which the heads belong to $\mathscr{X}$ but the tails do not, is referred to as the in-cut of $\mathscr{X}$, and is denoted by ${\partial}^{-}_{\mathscr{G}}{\mathscr{X}} \subset \mathscr{E}$. The cardinality of ${\partial}^{-}_{\mathscr{G}}{\mathscr{X}}$ is called the in-degree of $\mathscr{X}$, and is characterized as ${d}^{-}_{\mathscr{G}}{\mathscr{X}} =  |{\partial}^{-}_{\mathscr{G}}{\mathscr{X}}|$. Notice that by definition there are no parallel arcs in the graphical representation described above. In other words, if two edges share the same pair of head and tail, then they are identical. A matrix $W \in \mathbb{R}^{|\mathscr{V}|\times|\mathscr{V}|}$ is called an in-weighting on $\mathscr{G}$ if $\forall \{ \nu_i,\nu_j \} \subset \mathscr{V}, (\nu_i,\nu_j) \not\in \mathscr{E} \rightarrow \left[W\right]_{ji} = 0$. For a given digraph $\mathscr{G} = (\mathscr{V},\mathscr{E})$ and any pair of vertices $\{\nu_i,\nu_j\} \subset \mathscr{V}$, let edge $\epsilon := (\nu_i,\nu_j) \in \mathscr{E}\cup\{\epsilon\}$. The \emph{edge-index} of $\epsilon$ is defined as a ${|\mathscr{V}|\times|\mathscr{V}|}$ matrix with exactly one non-zero element which is a $1$ located at its $j-$th row and $i-$th column. This matrix is represented by $\Gamma(\epsilon) = \Gamma((\nu_i,\nu_j))$. Similarly, the \emph{vertex-index} of any $\nu_i \in \mathscr{V}$ is defined as a ${|\mathscr{V}| \times 1}$ column vector with exactly one non-zero element which is a $1$ located at its $i$-th row. This vector is denoted by $\boldsymbol{\sigma}(\nu_i)$. The adjacency matrix of $\mathscr{G}$ is given by $A(\mathscr{G}) = \sum_{\epsilon \in \mathscr{E}} \Gamma(\epsilon)$, its \emph{degree matrix} is defined as $\Delta(\mathscr{G}) = \sum_{\nu \in \mathscr{V}} {d}^{-}_{\mathscr{G}}\{\nu\}\Gamma((\nu,\nu))$, and the corresponding in-degree graph Laplacian is given by $\mathscr{L}(\mathscr{G}) = \Delta(\mathscr{G}) - A(\mathscr{G})$.

Given an integer $k \in \mathbb{N}$, a set of (possibly repeated) indices $\{\alpha_1,\alpha_2,\ldots,\alpha_k\}$ $ \subseteq$ $\mathbb{N}_{|\mathscr{V}|}$ and two vertices $\tau,\nu $ $\in$ $ \mathscr{V}$, an ordered sequence of edges of the form $\mathscr{W}$ $:=$ $({\tau},{\nu}_{\alpha_1})$ $,$ $({\nu}_{\alpha_1},{\nu}_{\alpha_2})$ $,$ $ \ldots$ $,$ $({\nu}_{\alpha_{k-1}},{\nu}_{\alpha_{k}})$ $,$ $({\nu}_{\alpha_{k}},{\nu})$ is called a \emph{${\tau}{\nu}$ walk} with start-node $\tau$, end-node $\nu$ and \emph{length} $k+1$. A cycle on node $\nu$ refers to a $\tau\nu$ walk where $\tau = \nu$. If $W$ is an in-weighting on $\mathscr{G}$, then $\omega(\mathscr{W},W) = \prod _{(\nu_i,\nu_j)\in\mathscr{W}} \left[W\right]_{ij}$ is referred to as the weight of walk $\mathscr{W}$ w.r.t $W$. In the same venue, the number of ${\tau}{\nu}$ walks with length $k$, denoted by $\textrm{c}_k(\mathscr{G};{\tau},{\nu})$, is called the \emph{$k-$th connectivity} of $\tau$ to $\nu$ in digraph $\mathscr{G}$; and by convention, $\textrm{c}_0(\mathscr{G};{\tau},{\nu}) = 0$ if $\tau \neq \nu$, while $\textrm{c}_0(\mathscr{G};{\nu},{\nu}) = 1$. Moreover, the integer 
\begin{align}
&\textrm{d}(\mathscr{G};{\nu_j},{\nu_i}) = \min_{k\in\mathbb{W},\textrm{c}_k({\nu_j},{\nu_i}) \neq 0}\{k\},
\nonumber
\end{align} is referred to as the distance from ${\nu_j}$ to ${\nu_i}$ in $\mathscr{G}$, and by convention $\textrm{d}(\mathscr{G};{\nu_j},{\nu_i}) = \infty$ if $\forall k\in\mathbb{N}, \textrm{c}_k(\mathscr{G};{\nu_j},{\nu_i}) = 0$. For any $\{{s},{p}\} \subset \mathbb{N}_{|\mathscr{V}|}$, $\Omega^{k}(\mathscr{G};{\nu_s,\nu_p})$ is the set of all ${\nu_s}{\nu_p}$ walks in $\mathscr{G}$ with length $k$. Similarly for $\{{s},{i},{p}\}$ $\subset$ $\mathbb{N}_{|\mathscr{V}|},$ $\Omega^{k}(\mathscr{G};{\nu_s,\nu_i,\nu_p})$ $=$ $\{ \mathscr{W}$ $\in$ $\Omega^{k}(\mathscr{G};{\nu_s,\nu_k}); (\nu_s,\nu_i)$ $\in$ $\mathscr{W} \}$, i.e. the set of $\nu_s \nu_k$ walks that include the edge $(\nu_s,\nu_i)$. Given a set of walks $\Omega$ in digraph $\mathscr{G}$ and in-weightings $W_1$ and $W_2$ on $\mathscr{G}$, the functions
\begin{align} 
& \Phi(\Omega, W_1) = \sum_{\mathscr{W}\in\Omega} \omega(\mathscr{W},W_1), \nonumber \\
& \Psi(\Omega, W_1, W_2) = \Phi(\Omega, W_1) - \Phi(\Omega, W_2), \nonumber
\end{align}
are defied, which will find use in the proof of the main theorem in Section~\ref{sec:derivatives} that follows. It is also known for an in-weighting $W$ on $\mathscr{G}$, and vertices $\{\nu_s,\nu_p\} \subset \mathscr{V}$, that \cite{BiggsGraphTheory}:
\begin{align}
&\Phi(\Omega^{k}(\mathscr{G};{\nu_s,\nu_p}),W) = \left[W^{k}\right]_{ps}.
\label{eq:numWalks}
\end{align}

\section{Derivatives of the Consensus Response}\label{sec:derivatives}

Given $n \in \mathbb{N}$, consider a multi-agent system comprised of a set $\mathscr{S}$ $ = $ $\{ x_i, i \in \mathbb{N}_n \}$ of $n$ single integrator agents, where $x_i$, $i \in \mathbb{N}_n$ is the scalar state of agent $i$. Under the linear agreement protocol, if the interaction structure between the agents is represented by a directed information flow graph $\mathscr{G} = (\mathscr{V},\mathscr{E})$, where $\mathscr{V}$ $=$ $\{ \nu_i, i \in \mathbb{N}_n \}$ and $\forall i \in \mathbb{N}_n$, $\nu_i$ corresponds to $x_i$, then the dynamic evolution law for the agents is given by:
\begin{equation}
\dot{\mathbf{x}}(t) = -\mathscr{L}(\mathscr{G})\mathbf{x}(t), \; t > 0, 
\label{eq:laplaciandynamics}
\end{equation} where $\mathbf{x}(t) = \left(x_1(t),x_2(t),\ldots,x_n(t)\right)^{T}$. For an initial condition $\mathbf{x}(0) \in \mathbb{R}^{|\mathscr{V}|}$, the matrix exponential solution to \eqref{eq:laplaciandynamics} can be derived as:
\begin{equation}
\mathbf{x}(t) = e^{-\mathscr{L}(\mathscr{G})t}\mathbf{x}(0), \; t \geqslant 0,
\label{eq:matrixsolutionlaplaciandynamics}
\end{equation} and for a particular agent $x_i \in \mathscr{S}$ represented by the vertex $\nu_i \in \mathscr{V}$, the temporal evolution of its state is then given by:
\begin{equation}
{x_i}(t) = \boldsymbol{\sigma}(\nu_i)^{T}e^{-\mathscr{L}(\mathscr{G})t}\mathbf{x}(0),  \; t \geqslant 0.
\label{eq:agentlaplaciandynamics}
\end{equation}

The next theorem paves the way for a method to detect and isolate the failure of any links in the network and at the same instant as they fail. The latter's significance is better understood upon noting that if the time of failure is random and has a continuous sample space, then  ``simultaneous'' failure of more than one link is a measure zero event, hence justifying the focus of investigation in this paper, which is on the ``single" link failures. It is further assumed that at each instant of time, the designer is given access to the response of a subset of agents as well as the network information flow digraph prior to the failure. In the case of detection, the designer is interested in determining the existence of any single link failure in the network at the instant of failure. For the isolation problem, however, the designer would like to determine ``instantaneously", not only the existence of a failure, but also its location. That is to determine which link, if any, has failed and exactly at the same instant as it fails.

The proposed method is based on the derivatives of the consensus response given in \eqref{eq:matrixsolutionlaplaciandynamics}. The proof ingredients are as follows. Terms of the form $(-\mathscr{L}(\mathscr{G}))^k$ appear upon taking the $k-$th derivative of \eqref{eq:matrixsolutionlaplaciandynamics}. Corresponding to a digraph $\mathscr{G}$, a new digraph $\tilde{\mathscr{G}}$ is defined by adding a self-loop on each node and it is then noted that $-\mathscr{L}(\mathscr{G})$ define an in-weighting on $\tilde{\mathscr{G}}$. Thence, the stage is set for the application of the summation formula given in \eqref{eq:numWalks} and the rest of the proof carries through by partitioning the set of walks over which \eqref{eq:numWalks} is summed.

In the theorem, $\mathscr{G}_1$ represents the original (healthy system) digraph and $\mathscr{G}_2$ is the digraph that is missing a single link. The removed link is $\epsilon := (\nu_i,\nu_j)$ and agent $x$ corresponding to vertex $\nu_p$ is an agent whose response $x(t)$ is being observed by the designer.

\begin{figure*}[!t]
\normalsize
\setcounter{MYtempeqncnt}{\value{equation}}
\setcounter{equation}{4}
\begin{equation}
\nabla(k) = \sum_{s=1}^{|\mathscr{V}|}[\Phi(\Omega^{k}(\tilde{\mathscr{G}}_1;{\nu_s,\nu_p}),-\mathscr{L}(\mathscr{G}_1)) - \Phi(\Omega^{k}(\tilde{\mathscr{G}}_2;{\nu_s,\nu_p}),-\mathscr{L}(\mathscr{G}_2))]{x}_s(0),
\label{eq:sumWeights}
\end{equation}
\begin{equation}
\nabla(k) =  \sum_{s=1}^{|\mathscr{V}|}\Psi(\Omega^{k}(\tilde{\mathscr{G}}_2;{\nu_s,\nu_p}),-\mathscr{L}(\mathscr{G}_1),-\mathscr{L}(\mathscr{G}_2)){x}_s(0).
\label{eq:sumWeightsoneSet}
\end{equation}
\begin{equation}
{\nabla}_1(k) =  \sum_{s=1}^{|\mathscr{V}|}\Psi(\Omega^{k}(\tilde{\mathscr{G}}_2;{\nu_s,\nu_p})\fgebackslash\Omega^{k}(\tilde{\mathscr{G}}_2;{\nu_s,\nu_i,\nu_p}),-\mathscr{L}(\mathscr{G}_1),-\mathscr{L}(\mathscr{G}_2)){x}_s(0).
\label{eq:sumWeightspartitioned1}
\end{equation}
\begin{equation}
{\nabla}_2(k) =   \sum_{s=1}^{|\mathscr{V}|} \Psi(\Omega^{k}(\tilde{\mathscr{G}}_2;{\nu_s,\nu_i,\nu_p}),-\mathscr{L}(\mathscr{G}_1),-\mathscr{L}(\mathscr{G}_2)){x}_s(0).
\label{eq:sumWeightspartitioned2} 
\end{equation}
\begin{align}
\nabla_i(k) &= [\Phi(\Omega^{k}(\tilde{\mathscr{G}}_1;{\nu_i,\nu_p}),-\mathscr{L}(\mathscr{G}_1)) - \Phi(\Omega^{k}(\tilde{\mathscr{G}}_2;{\nu_i,\nu_p}),-\mathscr{L}(\mathscr{G}_2))]{x}_i(0) \nonumber \\
& \overset{a}{=} [\Phi(\Omega^{k}(\tilde{\mathscr{G}}_1;{\nu_i,\nu_p})\fgebackslash\Omega^{k}(\tilde{\mathscr{G}}_1;{\nu_i,\nu_i,\nu_p}),-\mathscr{L}(\mathscr{G}_1)) + \Phi(\Omega^{k}(\tilde{\mathscr{G}}_1;{\nu_i,\nu_i,\nu_p}),-\mathscr{L}(\mathscr{G}_1)) - \nonumber \\ & \; \;\; \; \:  \Phi(\Omega^{k}(\tilde{\mathscr{G}}_2;{\nu_i,\nu_p})\fgebackslash\Omega^{k}(\tilde{\mathscr{G}}_2;{\nu_i,\nu_i,\nu_p}),-\mathscr{L}(\mathscr{G}_2)) - \Phi(\Omega^{k}(\tilde{\mathscr{G}}_2;{\nu_i,\nu_i,\nu_p}),-\mathscr{L}(\mathscr{G}_2)) ]{x}_i(0) \nonumber \\
& \overset{b}{=} [\Phi(\Omega^{k}(\tilde{\mathscr{G}}_1;{\nu_i,\nu_i,\nu_p}),-\mathscr{L}(\mathscr{G}_1)) -  \Phi(\Omega^{k}(\tilde{\mathscr{G}}_2;{\nu_i,\nu_i,\nu_p}),-\mathscr{L}(\mathscr{G}_2)) ]{x}_i(0) \nonumber \\
& = \{ \left[-\mathscr{L}({\mathscr{G}_1})\right]_{ii} \textrm{c}_{k-1}(\mathscr{G}_1;{\nu}_i,{\nu}_p) -  \left[-\mathscr{L}({\mathscr{G}_2})\right]_{ii} \textrm{c}_{k-1}(\mathscr{G}_2;{\nu}_i,{\nu}_p) \} {x}_i(0) \nonumber \\& = [(-{d}^{-}_{\mathscr{G}_1}\{\nu_i\})\textrm{c}_{k-1}(\mathscr{G}_1;{\nu}_i,{\nu}_p) -  (-{d}^{-}_{\mathscr{G}_1}\{\nu_i\} + 1)\textrm{c}_{k-1}(\mathscr{G}_1;{\nu}_i,{\nu}_p) ]{x}_i(0) = -\textrm{c}_{k-1}(\mathscr{G}_1;{\nu}_i,{\nu}_p){x}_i(0).
\label{eq:sumWeightsIJ_I}
\end{align}
\begin{align}
\nabla_j(k) & = [ \Phi(\Omega^{k}(\tilde{\mathscr{G}}_1;{\nu_j,\nu_p}),-\mathscr{L}(\mathscr{G}_1)) - \Phi(\Omega^{k}(\tilde{\mathscr{G}}_2;{\nu_j,\nu_p}),-\mathscr{L}(\mathscr{G}_2)) ] {x}_j(0) \nonumber \\
& \overset{1}{=} [\Phi(\Omega^{k}(\tilde{\mathscr{G}}_1;{\nu_j,\nu_p}),-\mathscr{L}(\mathscr{G}_1)) - \Phi(\Omega^{k}(\tilde{\mathscr{G}}_1;{\nu_j,\nu_p}) \fgebackslash \Omega^{k}(\tilde{\mathscr{G}}_1;{\nu_j,\nu_i,\nu_p}),-\mathscr{L}(\mathscr{G}_2)) ] {x}_j(0) \nonumber \\
& \overset{2}{=} [\Phi(\Omega^{k}(\tilde{\mathscr{G}}_1;{\nu_j,\nu_p}),-\mathscr{L}(\mathscr{G}_1)) - \Phi(\Omega^{k}(\tilde{\mathscr{G}}_1;{\nu_j,\nu_p}) \fgebackslash \Omega^{k}(\tilde{\mathscr{G}}_1;{\nu_j,\nu_i,\nu_p}),-\mathscr{L}(\mathscr{G}_1)) ] {x}_j(0) \nonumber \\
& = \Phi(\Omega^{k}(\tilde{\mathscr{G}}_1;{\nu_j,\nu_i,\nu_p}),-\mathscr{L}(\mathscr{G}_1))  {x}_j(0) \overset{3}{=} \left[-\mathscr{L}({\mathscr{G}_1})\right]_{ij}\textrm{c}_{k-1}(\mathscr{G}_1;{\nu}_i,{\nu}_p){x}_j(0) = \textrm{c}_{k-1}(\mathscr{G}_1;{\nu}_i,{\nu}_p){x}_j(0).
\label{eq:sumWeightsIJ_J}
\end{align}
\begin{equation}
\bar{\nabla}(k) = \sum_{s \in \mathbb{N}_{|\mathscr{V}|}\fgebackslash\{i,j\}} [ \Phi(\Omega^{k}(\tilde{\mathscr{G}}_1;{\nu_s,\nu_p}),-\mathscr{L}(\mathscr{G}_1)) - \Phi(\Omega^{k}(\tilde{\mathscr{G}}_2;{\nu_s,\nu_p}),-\mathscr{L}(\mathscr{G}_2))]{x}_s(0) \overset{\alpha}{=} 0.
\label{eq:sumWeightsIJ_BAR}
\end{equation}
\setcounter{equation}{\value{MYtempeqncnt}}
\hrulefill
\vspace*{4pt}
\end{figure*}
\addtocounter{equation}{9}

\begin{mytheorem}\label{theo:detection1} Given a multi-agent system $\mathscr{S}$ and its associated digraph $\mathscr{G}_1 = (\mathscr{V},\mathscr{E}_1)$, consider a vertex ${\nu_p} \in \mathscr{V}$ corresponding to agent $x \in \mathscr{S}$, and an edge $\epsilon:=(\nu_j,\nu_i)\in\mathscr{E}_1$, and denote $\mathscr{G}_2 = (\mathscr{V},\mathscr{E}_1\fgebackslash\{\epsilon\})$. Starting from the same initial condition $\mathbf{x}(0)$, for $t\geqslant0$, let $ x_{\mathscr{G}_1}(t)$ and $x_{\mathscr{G}_2}(t)$ denote the state of the agent $x$ calculated in the digraphs $\mathscr{G}_1$ and $\mathscr{G}_2$, respectively; and define $\nabla(k) := \frac{ {\textrm{\emph{d}}}^{k} } {\textrm{\emph{d}} t^{k}}(x_{\mathscr{G}_1})(0^{+})-\frac{ {\textrm{\emph{d}}}^{k} } {\textrm{\emph{d}} t^{k}}(x_{\mathscr{G}_2})(0^{+})$. The following statements hold true: $(i)$ $\forall k \leqslant \textrm{\emph{d}}(\mathscr{G}_1;{\nu}_j,{\nu_p}) - 1, \nabla(k) = 0$, and $(ii)$ for $k$ $=$ $\mbox{\emph{d}}(\mathscr{G}_1;{\nu}_j,{\nu_p}),$ $\nabla(k) = \textrm{c}_{k-1}(\mathscr{G}_1;{\nu}_i,{\nu_p})(x_j(0) - x_i(0))$.
\end{mytheorem}

\textbf{Proof.} Define $\tilde{\mathscr{G}}_1$ $=$ $(\mathscr{V},\mathscr{E}_1 \cup_{\nu\in\mathscr{V}}\{(\nu,\nu)\})$ and $\tilde{\mathscr{G}}_2$ $=$ $(\mathscr{V},\mathscr{E}_2 \cup_{\nu\in\mathscr{V}}\{(\nu,\nu)\})$ and note that $-\mathscr{L}(\mathscr{G}_1)$ and $-\mathscr{L}(\mathscr{G}_2)$ define proper in-weightings on $\tilde{\mathscr{G}}_1$ and $\tilde{\mathscr{G}}_2$, respectively. Differentiating both sides of \eqref{eq:agentlaplaciandynamics} $k$ times yields: 
\begin{align}
\nabla(k) = \boldsymbol{\sigma}(\nu_p)^{T}((-\mathscr{L}(\mathscr{G}_1))^k - (-\mathscr{L}(\mathscr{G}_2))^k) \mathbf{x}(0),
\nonumber 
\end{align}
which can be rewritten as: 
\begin{align}
\nabla(k) = \sum_{s=1}^{|\mathscr{V}|}\left( \left[(-\mathscr{L}(\mathscr{G}_1))^{k}\right]_{ps} - \left[(-\mathscr{L}(\mathscr{G}_2))^{k}\right]_{ps}\right){x}_s(0),
\nonumber
\end{align}where ${x}_s(0)$ is a scalar value that denotes the initial state of the agent corresponding to vertex $\nu_s \in \mathscr{V}$. Replacing from \eqref{eq:numWalks} into the preceding expression of $\nabla(k)$ leads to \eqref{eq:sumWeights} at the top of the succeeding page. 

For part $(i)$, note that if $k$ $\leqslant$ $\mbox{{d}}(\mathscr{G}_1;{\nu}_j,{\nu_p})$ $-$ $1$, then there are no walks of length $k$ that include $(\nu_j,\nu_i)$ as an edge and terminate at node $\nu_p$. Hence, for $k$ $\leqslant$ $\mbox{{d}}(\mathscr{G}_1;{\nu}_j,{\nu_p})$ $-$ $1$, $\Omega^{k}(\tilde{\mathscr{G}}_2;{\nu_s,\nu_p})$ $=$ $\Omega^{k}(\tilde{\mathscr{G}}_1;{\nu_s,\nu_p})$, since $\tilde{\mathscr{G}}_1$ and $\tilde{\mathscr{G}}_2$ differ only at edge $(\nu_j,\nu_i)$ and thus they have the same set of walks with no $(\nu_j,\nu_i)$ edge. Given that $\Omega^{k}(\tilde{\mathscr{G}}_2;{\nu_s,\nu_p})$ $=$ $\Omega^{k}(\tilde{\mathscr{G}}_1;{\nu_s,\nu_p})$, \eqref{eq:sumWeights} can be rewritten as \eqref{eq:sumWeightsoneSet} at the top of the next page. The next step is to partition the set of walks $\Omega^{k}(\tilde{\mathscr{G}}_2;{\nu_s,\nu_p})$ into two disjoint sets, $\Omega^{k}(\tilde{\mathscr{G}}_2;{\nu_s,\nu_p})$ and $\Omega^{k}(\tilde{\mathscr{G}}_2;{\nu_s,\nu_p})$ $\fgebackslash$ $\Omega^{k}(\tilde{\mathscr{G}}_2;{\nu_s,\nu_i,\nu_p})$. Thence, $\nabla(k)$ in \eqref{eq:sumWeightsoneSet} can be rewritten as $\nabla(k)$ $=$ $\nabla_1(k)$ $+$ $\nabla_2(k)$, where ${\nabla}_1(k)$ and ${\nabla}_2(k)$ are given by \eqref{eq:sumWeightspartitioned1} and \eqref{eq:sumWeightspartitioned2}, and they correspond to the contributions made by the walks in $\Omega^{k}(\tilde{\mathscr{G}}_2;{\nu_s,\nu_p})$ $\fgebackslash$ $\Omega^{k}(\tilde{\mathscr{G}}_2;{\nu_s,\nu_i,\nu_p})$ and $\Omega^{k}(\tilde{\mathscr{G}}_2;{\nu_s,\nu_i,\nu_p})$, respectively. To finish the proof of part $(i)$, note that $\forall \mathscr{W}$ $\in$ $\Omega^{k}(\tilde{\mathscr{G}}_2;{\nu_s,\nu_p})$ $\fgebackslash$ $\Omega^{k}(\tilde{\mathscr{G}}_2;{\nu_s,\nu_i,\nu_p})$, $\omega(\mathscr{W},-\mathscr{L}(\mathscr{G}_1))$ $=$ $\omega(\mathscr{W},-\mathscr{L}(\mathscr{G}_2))$, since any such walk $\mathscr{W}$ includes neither of the edges $(\nu_i,\nu_i)$ and $(\nu_j,\nu_i)$, which are the only edges at which $-\mathscr{L}(\mathscr{G}_1)$ and $-\mathscr{L}(\mathscr{G}_2)$ differ. It is therefore true that $\forall s \in \mathbb{N}_{|\mathscr{V}|},$ $\Psi(\Omega^{k}(\tilde{\mathscr{G}}_2;{\nu_s,\nu_p})$ $\fgebackslash$ $\Omega^{k}(\tilde{\mathscr{G}}_2;{\nu_s,\nu_i,\nu_p}),$ $-\mathscr{L}(\mathscr{G}_1),$ $-\mathscr{L}(\mathscr{G}_2))$ $=$ $0$, and ${\nabla}_1(k)$ $=$ $0$. On the other hand, from $m$ $\leqslant$ $\mbox{{d}}(\mathscr{G}_1;{\nu}_j,{\nu_p})$ $-1$ and $(\nu_j,\nu_i)$ $ \in$ $ \mathscr{E}_1$, it follows that $m$ $\leqslant$ $\mbox{{d}}(\mathscr{G}_1;{\nu}_i,{\nu_p})$ $\leqslant$ $\mbox{{d}}(\mathscr{G}_1;{\nu}_j,{\nu_p})$ $-1$, which in turn implies that $\Omega^{k}(\tilde{\mathscr{G}}_2;{\nu_s,\nu_i,\nu_p})$ $=$ $\varnothing$. Because with $m$ $\leqslant$ $\mbox{{d}}(\mathscr{G}_1;{\nu}_i,{\nu_p})$ there can be no $\nu_s\nu_p$ walks of length $k$ in $\tilde{\mathscr{G}}_2$ with $(\nu_s,\nu_i)$ as an edge, for otherwise one can remove $(\nu_s,\nu_i)$ and construct a $\nu_i\nu_p$ walk in $\mathscr{G}_1$, whose length is strictly less than $\mbox{{d}}(\mathscr{G}_1;{\nu}_i,{\nu_p})$. Now from $\Omega^{k}(\tilde{\mathscr{G}}_2;{\nu_s,\nu_i,\nu_p})$ $=$ $\varnothing$ it follows that $\forall s$ $ \in$ $ \mathbb{N}_{|\mathscr{V}|},$ $\Psi(\Omega^{k}(\tilde{\mathscr{G}}_2;{\nu_s,\nu_i,\nu_p}),$ $-\mathscr{L}(\mathscr{G}_1),$ $-\mathscr{L}(\mathscr{G}_2))$ $ = 0$ and ${\nabla}_2(m)$ $=$ $0$, as well. Thus far, it is shown that ${\nabla}_1(k)$ $=$ ${\nabla}_2(k)$ $=$ $0$, whence $\nabla(k)$ $=$ $\nabla_1(k)$ $+$ $\nabla_2(k)$ $=$ $0$, completing the proof for $k$ $\leqslant$ $\mbox{{d}}(\mathscr{G}_1;{\nu}_j,{\nu_p})-1$.

For the case of $k$ $=$ ${\textrm{d}}(\mathscr{G}_1;{\nu}_j,{\nu_p})$ in part $(ii)$, first note that by conditioning on the choice of the start-node $s$, the summation in \eqref{eq:sumWeights} can be can be rewritten as: $\nabla(k)$ $=$ $\nabla_i(k)$ $+$ $\nabla_j(k)$ $+$ $\bar{\nabla}(k)$, where the three terms are defined in \eqref{eq:sumWeightsIJ_I} to \eqref{eq:sumWeightsIJ_BAR} at the top, and they measure the contributions due to the walks starting from nodes $\nu_i$, $\nu_j$, and the rest of the nodes, respectively. In the following paragraphs, the expressions for each of the the above three terms are simplified in the respective order, leading to the equation in part $(ii)$.

The first step in simplifying the expression of $\nabla_i(k)$ is $(\overset{a}{=})$, which follows by partitioning the sets $\Omega^{k}(\tilde{\mathscr{G}}_1;{\nu_i,\nu_p})$ and $\Omega^{k}(\tilde{\mathscr{G}}_2;{\nu_i,\nu_p})$. Next note that no $\nu_i\nu_p$ of length $m = \mbox{{d}}(\mathscr{G}_1;\nu_j,\nu_p)$ can include $(\nu_j,\nu_i)$ as an edge. This statement is trivial in the case of $\tilde{\mathscr{G}}_2$. To see why it is true for $\tilde{\mathscr{G}}_1$ as well, suppose to the contrary that there exists a $\nu_i\nu_p$ walk $\mathscr{W}_1$ of length $k$ with $(\nu_j,\nu_i)$ as an edge. There is therefore a cycle on $\nu_i$ whose length is at least $2$, and removing that from $\mathscr{W}_1$ yields a new $\nu_i\nu_p$ walk $\mathscr{W}_2$ with length at most $m-2$. Now $(\nu_j,\nu_i)\mathscr{W}_2$ is a $\nu_j\nu_p$ walk of length at most $m-1$, which is a contraction, since $k = d(\mathscr{G}_1;\nu_j,\nu_i)$.  The next step $(\overset{b}{=})$ follows upon the realization that $\Phi(\Omega^{k}(\tilde{\mathscr{G}}_1;{\nu_i,\nu_p}) $ $\fgebackslash$ $\Omega^{k}(\tilde{\mathscr{G}}_1;{\nu_i,\nu_i,\nu_p}),$ $-\mathscr{L}(\mathscr{G}_1))$ $=$ $\Phi(\Omega^{k}(\tilde{\mathscr{G}}_2;{\nu_i,\nu_p})$ $\fgebackslash$ $\Omega^{k}(\tilde{\mathscr{G}}_2;{\nu_i,\nu_i,\nu_p}),$ $-\mathscr{L}(\mathscr{G}_2))$, which is true since none of the walks involved include any of the edges $(\nu_j,\nu_i)$ or $(\nu_i,\nu_i)$. The rest of the equalities in \eqref{eq:sumWeightsIJ_I} follow through as a consequence of $\mbox{{d}}(\tilde{\mathscr{G}}_1;{\nu}_i,{\nu_p})$ $=$ $\mbox{{d}}(\tilde{\mathscr{G}}_2;{\nu}_i,{\nu_p})$ $\geqslant$ $k-1$, which implies that $(\nu_i,\nu_i)$ is the one and only self-loop that is included in any and all of the walks in $\Omega^{k}(\tilde{\mathscr{G}}_1;{\nu_i,\nu_i,\nu_p})$ or $\Omega^{k}(\tilde{\mathscr{G}}_2;{\nu_i,\nu_i,\nu_p})$. 

To simplify the expression for $\nabla_j(k)$, note that since $\tilde{\mathscr{G}}_2$ is derived by removing edge $(\nu_i,\nu_j)$, it holds true that $\Omega^{k}(\tilde{\mathscr{G}}_2;{\nu_j,\nu_p}) $ $=$ $ \Omega^{k}(\tilde{\mathscr{G}}_1;{\nu_j,\nu_p})$ $\fgebackslash$ $\Omega^{k}(\tilde{\mathscr{G}}_1;{\nu_j,\nu_i,\nu_p})$, which leads to $(\overset{1}{=})$ in \eqref{eq:sumWeightsIJ_J}. Next, since $k$ $=$ $\mbox{{d}}(\mathscr{G}_1;{\nu}_j,{\nu_p})$ none of the walks in $\Omega^{k}(\tilde{\mathscr{G}}_1;{\nu_j,\nu_p})$ include any self-loops. In particular, any walk $\mathscr{W}$ $\in$ $\Omega^{k}(\tilde{\mathscr{G}}_1;{\nu_j,\nu_p})$ $\fgebackslash$ $\Omega^{k}(\tilde{\mathscr{G}}_1;{\nu_j,\nu_i,\nu_p})$ includes neither of the edges $(\nu_i,\nu_i)$ and $(\nu_j,\nu_i)$, thence $\omega(\mathscr{W},-\mathscr{L}(\mathscr{G}_1))$ $=$ $\omega(\mathscr{W},-\mathscr{L}(\mathscr{G}_2))$, implying $(\overset{2}{=})$. To see $(\overset{3}{=})$, again bear in mind that $k$ $=$ ${\textrm{d}}(\mathscr{G}_1;{\nu}_j,{\nu_p})$, together with $(\nu_j,\nu_i)$ $\in$ $\mathscr{E}_1$, implies that ${\textrm{d}}(\mathscr{G}_1;{\nu}_i,{\nu_p})$ $\geqslant$ $k-1$ so that none of the walks in $\Omega^{k}(\tilde{\mathscr{G}}_1;{\nu_j,\nu_i,\nu_p})$ include any self-loops.

Finally, if $s$ $\not\in$ $\{i,j\}$, then $\forall \mathscr{W}$ $\in$ $\Omega^{k}(\tilde{\mathscr{G}}_1;{\nu_s,\nu_p})$ $\cup$ $\Omega^{k}(\tilde{\mathscr{G}}_2;{\nu_s,\nu_p})$, $\{(\nu_j,\nu_i),(\nu_i,\nu_i)\}$ $\cap$ $\mathscr{W}$ $=$ $\varnothing$, since otherwise it is possible to construct a $\nu_j\nu_p$ walk in $\mathscr{G}_1$, whose length is strictly less than $k = \mbox{{d}}(\mathscr{G}_1;{\nu}_j,{\nu_p})$. Hence, by an argument similar to the one leading to ${\nabla}_1(k) = 0$ in the proof of part $(i)$, $\Omega^{k}(\tilde{\mathscr{G}}_1;{\nu_s,\nu_p})$ $=$ $\Omega^{k}(\tilde{\mathscr{G}}_2;{\nu_s,\nu_p})$ and $\forall \mathscr{W}$ $\in$ $\Omega^{k}(\tilde{\mathscr{G}}_1;{\nu_s,\nu_p})$ $=$ $\Omega^{k}(\tilde{\mathscr{G}}_2;{\nu_s,\nu_p}),$ $\omega(\mathscr{W},-\mathscr{L}(\mathscr{G}_1))$ $=$ $\omega(\mathscr{W},-\mathscr{L}(\mathscr{G}_2))$, which justifies $(\overset{\alpha}{=})$ in \eqref{eq:sumWeightsIJ_BAR}. Putting all three together yields that $\nabla(k)$ $=$  $\nabla_i(k)$ $+$ $\nabla_j(k)$ $+$ $\bar{\nabla}(k)$ $=$ $\textrm{c}_{k-1}(\mathscr{G}_1;{\nu}_i,{\nu_p})$ $(x_j(0) - x_i(0))$ and the proposition holds true for $k$ $=$ ${\textrm{d}}(\mathscr{G}_1;{\nu}_j,{\nu_p})$ as well. \hfill {\scriptsize $\blacksquare$}

The following important remark highlights the key requirements that need to be satisfied if Theorem~\ref{theo:detection1} is to be exploited for detection and isolation of link failures.

\begin{myremark}\label{rem:jumpsDerivatives} In Theorem~\ref{theo:detection1}, if $\mbox{\emph{d}}(\mathscr{G}_1;{\nu}_j,{\nu_p}) = k$, then $\mbox{\emph{d}}(\mathscr{G}_1;{\nu}_i,{\nu_p}) $ $\geqslant$ $k -1$. On the other hand, the quantity $\textrm{c}_{k-1}(\mathscr{G}_1;{\nu}_i,{\nu}_p)(x_j(0) - x_i(0))$ that appears in part $(ii)$ of Theorem~\ref{theo:detection1} is nonzero only if $(x_j(0) - x_i(0)) \neq 0$ and $\mbox{\emph{d}}(\mathscr{G}_1;{\nu}_i,{\nu_p}) $ $\leqslant$ $ k -1$ or in this particular case, $\mbox{\emph{d}}(\mathscr{G}_1;{\nu}_i,{\nu_p}) $ $=$ $ k-1$. Hence, if the existence of a jump discontinuity in the $k-$th derivative of the output response at node $\nu_p$ is to serve as the basis for a method to detect the failure of edge $(\nu_j,\nu_i)$ at time $t = 0$, then it should be true that $x_j(0) - x_i(0) \neq 0$ and $\mbox{\emph{d}}(\mathscr{G}_1;{\nu}_i,{\nu_p})$ $ = $ $k-1$ $<$ $\mbox{\emph{d}}(\mathscr{G}_1;{\nu}_j,{\nu_p})$ $=$ $k$. The latter inequality is in perfect agreement with the sufficient condition stated in Corollary 1 of \cite{FailureDetectionCDC2012}.
\end{myremark}

In the next section, the focus is shifted to the problem of sensor placement, such that in line with Remark \ref{rem:jumpsDerivatives}, by observing the jumps in the derivatives of the responses of a subset of nodes, the designer can detect and/or isolate the failure of any single link across the network.

\section{Placement of the Sensors}\label{sec:sensorpalcement}

Throughout this section, $\mathscr{S} = \{x_i,i\in\mathbb{N}_n\}$ is a multi-agent system, with $n \in \mathbb{N}$ a fixed integer. The digraph $\mathscr{G} = (\mathscr{V},\mathscr{E})$ is associated with $\mathscr{S}$, where $\mathscr{V}$ $=$ $\{ \nu_s, s \in \mathbb{N}_n \}$ and $\forall s \in \mathbb{N}_n$, $\nu_s$ corresponds to $x_s$. Moreover, $z \in \mathbb{N}_n$ is a fixed integer denoting the highest order of the derivatives of the responses of the consensus dynamics on $\mathscr{S}$, to which the designer has access. Similarly, with $\{k,p,i,j\}$ $\subset$ $\mathbb{N}_n$, the order of the derivatives is denoted by $k \in \mathbb{N}_{z}$, $\nu_p \in \mathscr{V}$ is a node whose response is being observed, $\epsilon := (\nu_j,\nu_i) \in \mathscr{E}$ is an edge whose failure at time $t = 0$ is to be detected and/or isolated  by the designer and it is further assumed that $x_i(0) \neq x_j(0)$, i.e. whenever an edge fails, the state values at its head and tail nodes do not coincide.

Given $k \in \mathbb{N}_z$, let $\mathscr{R}_{k}$ and $\mathscr{R}_{0}$ be binary relations between the vertices and edges of $\mathscr{G}$, such that $\forall \nu_p \in \mathscr{V}$ and $\forall \epsilon \in \mathscr{E}$, where $\epsilon = (\nu_j,\nu_i)$ for some $\{\nu_j,\nu_i\} \subset \mathscr{V}$, $(\nu_p,\epsilon) \in \mathscr{R}_{j} \leftrightarrow (\textrm{d}(\mathscr{G};{\nu_j},{\nu_p}) = k) \wedge (\textrm{d}(\mathscr{G};{\nu_i},{\nu_p}) = k-1)$ and $(\nu_p,\epsilon) \in \mathscr{R}_0 \leftrightarrow \forall k \in \mathbb{N}_z, (\nu_p,\epsilon) \not\in \mathscr{R}_{k}$. Note that with $\epsilon = (\nu_j,\nu_i)$, by definition $(\nu_i,\epsilon) \in \mathscr{R}_{1}$ and $(\nu_j,\epsilon) \in \mathscr{R}_{0}$.

The preceding definitions are motivated by the observation made in Remark \ref{rem:jumpsDerivatives} that if the existence of jumps in the $k-$th derivative of the response of a node $\nu_p$  is to serve as an indicator for the failure of edge $(\nu_j,\nu_i)$, then it should hold true that $(\textrm{d}(\mathscr{G};{\nu_j},{\nu_p}) = k) \wedge (\textrm{d}(\mathscr{G};{\nu_i},{\nu_p}) = k-1)$. The problems of detection and isolation are now posed and addressed as sensor placement problems in the following \cite{Dhillon03sensorplacement,Leskovec}.

\begin{myproblem}\label{prob:detection} \emph{[Detection]} Given a digraph $\mathscr{G} = (\mathscr{V},\mathscr{E})$, propose a (preferably minimal) subset of vertices $\mathscr{M}_D \subset \mathscr{V}$ such that $\forall \epsilon \in \mathscr{E}, \exists \nu_p \in \mathscr{M}_D, (\nu_p,\epsilon) \not\in \mathscr{R}_{0}$.
\end{myproblem}

Let $\mathscr{M}_D \subset \mathscr{V}$ be any solution to Problem \ref{prob:detection}, then a link has failed at (the arbitrarily shifted) time $t = 0$ if, and only if, there exists some $\nu_p \in \mathscr{M}_D$ and $k \in \mathbb{N}_z$ such that $\frac{ {\textrm{\emph{d}}}^{k} } {\textrm{\emph{d}} t^{k}}x_p(0^{+})-\frac{ {\textrm{\emph{d}}}^{k} } {\textrm{\emph{d}} t^{k}}x_p(0^{-}) \neq 0$. That is, by observing the first $z$ derivatives of the responses of all the nodes in $\mathscr{M}_D$, the designer determines that a failure occurs if, and only if, a jump is observed in one or more nodes in $\mathscr{M}_D$. For the isolation problem the following extra tool will come handy. Given a subset of nodes $\mathscr{M}$ and any edge $\epsilon \in \mathscr{E}$ define the \emph{indicator set} of $\epsilon$ w.r.t. $\mathscr{M}$ as the function $\mathscr{I}: \mathscr{P}(\mathscr{V}) \to \mathscr{P}({(\mathbb{N}_z\cup\{0\}) \times \mathscr{M}})$, given by $\mathscr{I}(\mathscr{M};{\epsilon}) = \{ (k,\nu_p) \in (\mathbb{N}_z\cup\{0\}) \times \mathscr{M}; (\nu_p,\epsilon) \in \mathscr{R}_{k} \}$.

\begin{myproblem}\label{prob:isolation} \emph{[Isolation]} Given a digraph $\mathscr{G} = (\mathscr{V},\mathscr{E})$, propose a (preferably minimal) subset of vertices $\mathscr{M}_I \subset \mathscr{V}$ such that $\mathscr{M}_I$ is a solution to Problem \ref{prob:detection} and $\forall \alpha, \beta \in \mathscr{E}, \alpha \neq \beta  \leftrightarrow \mathscr{I}(\mathscr{M}_I;{\alpha}) \neq \mathscr{I}(\mathscr{M}_I;{\beta})$.
\end{myproblem}

Given a solution $\mathscr{M}_I$ to Problem \ref{prob:isolation}, any edge $\epsilon$ is uniquely determined by the set of ordered pairs $\mathscr{I}(\mathscr{M};{\epsilon})$. Hence, the edge $\epsilon$ has failed at (the arbitrarily shifted) time $t = 0$ if, and only if, $\forall (k,\nu_p) \in \mathscr{I}(\mathscr{M};{\epsilon}),$ $k > 0,$ $\frac{{\textrm{\emph{d}}}^{k}}{\textrm{\emph{d}} t^{k}}x_p(0^{+})-\frac{{\textrm{\emph{d}}}^{k}}{\textrm{\emph{d}} t^{k}}x_p(0^{-}) \neq 0,$ and $\forall (0,\nu_p) \in \mathscr{I}(\mathscr{M};{\epsilon}),$ $\forall m \in \mathbb{N}_{z},$ $\frac{{\textrm{\emph{d}}^m}}{\textrm{\emph{d}} t^{m}}x_p(0^{+}) - \frac{{\textrm{\emph{d}}}^{m}}{\textrm{\emph{d}} t^{m}}x_p(0^{-}) = 0$. In other words, by observing the first $z$ derivatives of the responses of all the nodes $\nu_p$ in $\mathscr{M}_I$, the designer determines that the edge $\epsilon$ has failed if, and only if, a jump is observed in the $k-$th derivative of the response of any node $\nu_p$ for which $(k,\nu_p) \in \mathscr{I}(\mathscr{M};{\epsilon})$, $k > 0$, and no jumps is observed in the first $z$ derivatives of the responses of all nodes $\nu_p$ for which $(0,\nu_p) \in \mathscr{I}(\mathscr{M};{\epsilon})$. 


In the following subsections, efficient algorithms are proposed for the selection of the observation points, which are the subsets $\mathscr{M}_D$ and $\mathscr{M}_I$ satisfying the requirement of the problems \ref{prob:detection} and \ref{prob:isolation}.

\subsection{Detection of All Failures: Coverage}\label{sec:coverage}

For all $\mathscr{M} \subset \mathscr{V}$ define $f_D: \mathscr{P}\mathscr{V}) \to \mathscr{P}(\mathbb{N}_{|\mathscr{E}|}\cup\{0\})$, given by $f_D(\mathscr{M}) = |\{\epsilon \in \mathscr{E}; \forall \nu_p \in {\mathscr{M}}, (\nu_p,\epsilon) \in \mathscr{R}_{0}\}|$. Note that if $\{\hat{\mathscr{M}},\bar{\mathscr{M}}\} \subset \mathscr{P}(\mathscr{V})$ are such that $\hat{\mathscr{M}} \subset \bar{\mathscr{M}}$, then for all $\nu \in \mathscr{V}$ it holds true that $f_D(\hat{\mathscr{M}}\cup\{\nu\}) - f_D(\hat{\mathscr{M}}) \leqslant f_D(\bar{\mathscr{M}}\cup\{\nu\}) - f_D(\bar{\mathscr{M}})$, i.e. $-f_D(.)$ a submodular set function from $\mathscr{P}(\mathscr{V})$ to $\mathbb{N}_{|\mathscr{E}|}\cup\{0\}$, which satisfies the diminishing returns or discrete concavity property \cite{Leskovec}. Routine \ref{routine:detection} uses this property of the set function $f_D$ to compute a solution $M_D \subset \mathscr{V}$ to Problem \ref{prob:detection} using an efficient greedy heuristic.

\begin{algorithm}
\caption{Determine a Solution $\mathscr{M}_D$ to Problem \ref{prob:detection}}
\label{routine:detection}
\begin{algorithmic}[1]
\REQUIRE $\mathscr{G} = (\mathscr{V},\mathscr{E})$
\State $\mathscr{M}_D \Leftarrow \varnothing$
\WHILE{$f_D(\mathscr{M}_D) \neq 0$}
\STATE $\nu  \Leftarrow \arg\min \{ f_D({\mathscr{M}}_D\cup\{\nu\}) - f_D({\mathscr{M}}_D);{{\nu} \in \mathscr{V}\fgebackslash}{\mathscr{M}}_D\}$
\STATE ${\mathscr{M}}_D \Leftarrow {\mathscr{M}}_D\cup\{\nu\}$
\ENDWHILE
\ENSURE ${\mathscr{M}}_D$
\end{algorithmic}
\end{algorithm}

\begin{myremark}\label{rem:rout1} The function $f_D(\mathscr{M}_1)$ measures the coverage of set $\mathscr{M}_D$ by counting the number of links that are not yet covered by $\mathscr{M}_D$. At each iteration of Routine \ref{routine:detection}, the extra node $\nu$ is selected and added to $\mathscr{M}_D$ such that the number of newly covered links is maximized. Note that since for $\epsilon = (\nu_j,\nu_i)$, $(\nu_i,\epsilon) \in \mathscr{R}_{1}$ it follows that $f_D(\mathscr{V}) = 0$, whence Routine \ref{routine:detection} is guaranteed to terminate. 
\end{myremark}

\subsection{Isolation of All Failures: Resolution}\label{sec:resolution}

Similarly to the previous section, for all $\mathscr{M} \subset \mathscr{V}$ define $f_I: \mathscr{P}(\mathscr{V}) \to \mathbb{N}_{|\mathscr{E}|}\cup\{0\}$, given by $f_I(\mathscr{M}) = |\{\epsilon \in \mathscr{E}; \exists \hat{\epsilon} \in\mathscr{E}\fgebackslash\{\epsilon\}, \mathscr{I}(\mathscr{M};{\epsilon}) = \mathscr{I}(\mathscr{M};{\hat{\epsilon}})\}|$, and note that $-f_I(.)$ is also submodular. Thus, a greedy heuristic similar to the one in Subsection \ref{sec:coverage} can be used to compute a solution $M_I \subset \mathscr{V}$ to Problem \ref{prob:isolation}, as follows.

\begin{algorithm}
\caption{Determine a Solution $\mathscr{M}_I$ to Problem \ref{prob:isolation}}
\label{routine:isolation}
\begin{algorithmic}[1]
\REQUIRE $\mathscr{G} = (\mathscr{V},\mathscr{E})$
\State $\mathscr{M}_I \Leftarrow \varnothing$
\WHILE {$f_I(\mathscr{M}_I) \neq 0 \And \mathscr{M}_I \neq \mathscr{V}$}
\STATE $\nu  \Leftarrow \arg\min \{ f_I({\mathscr{M}}_I\cup\{\nu\}) - f_I({\mathscr{M}}_I);{{\nu} \in \mathscr{V}\fgebackslash}{\mathscr{M}}_I\}$
\STATE ${\mathscr{M}}_I \Leftarrow {\mathscr{M}}_I\cup\{\nu\}$
\ENDWHILE
\IF{$f_I(\mathscr{M}_I) \neq 0$} \State $\mathscr{M}_I \Leftarrow \varnothing$
\ENDIF 
\ENSURE ${\mathscr{M}}_I$
\end{algorithmic}
\end{algorithm}

\begin{myremark}\label{rem:rout2} The function $f_I(\mathscr{M}_I)$ measures the resolution of set $\mathscr{M}_I$ by counting the number of links that are not uniquely identified through their relations $\mathscr{R}_{k}$ with the vertices of set $\mathscr{M}_I$. At each iteration of Routine \ref{routine:isolation}, the extra node $\nu$ is selected and added to $\mathscr{M}_I$ such that the resultant improvement in the resolution of $\mathscr{M}_I$ is maximized. Note that unlike Problem \ref{prob:detection}, it is possible for Problem \ref{prob:isolation} to have no solutions at all, in which case Routine \ref{routine:isolation} returns $\varnothing$. This occurs if and only if $f_I(\mathscr{V}) \neq 0$.
\end{myremark}

\section{Examples and Discussions}\label{sec:examples}

In the sequel, $\mathscr{V} = \{\nu_i,i\in \mathbb{N}_5\}$ is a set of five vertices, where for all $i\in \mathbb{N}_5$, vertex $\nu_i$ corresponds to agent $x_i$ in a multi-agent system $\mathscr{S} = \{x_i,i\in \mathbb{N}_5\}$. It is further assumed that if a link $(\nu_i,\nu_j), \{i,j\}\subset\mathbb{N}_5$ fails at $t = t_f$, then $x_i(t_f) \neq x_j(t_f)$.

\begin{figure}[ht]
\centering
\subfigure[]{
\includegraphics[width=32mm]{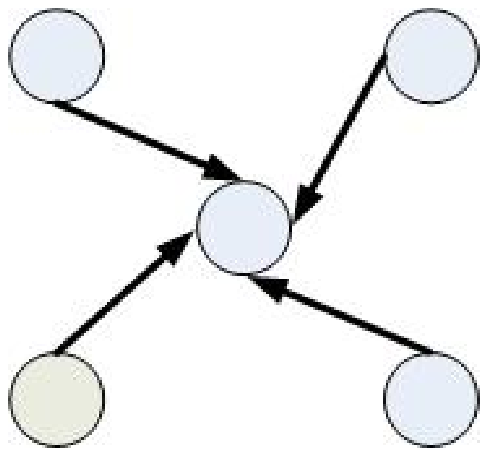}
\label{fig:star}
}
\subfigure[]{
\includegraphics[width=41mm]{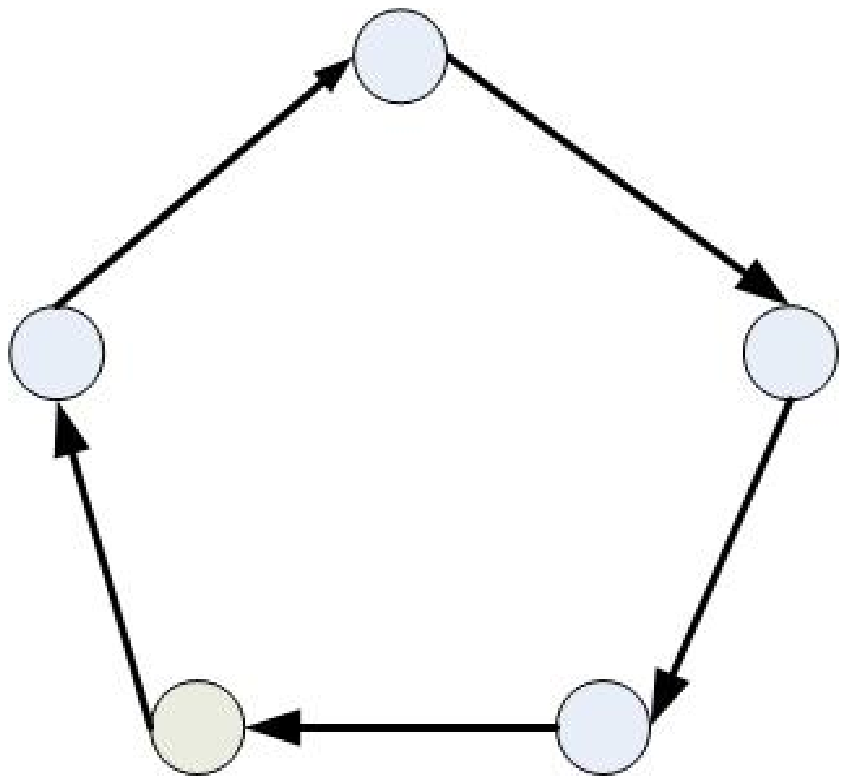}
\label{fig:cycle}
}
\label{fig:digraphs}
\caption{  A start digraph of size five and a directed cycle of length five are depicted in \subref{fig:star} and \subref{fig:cycle}, respectively.}
\end{figure}


As the first example, consider the case of a start network (Fig. \ref{fig:star}), where there are four edges in the network and all of them share the same head vertex $\nu_5$. In this case, the designer can detect the failure of any single edge in the network by observing the first derivative of the response of $x_5$; however, there are no subset of nodes that can be observed for isolating the failed edge. In fact $f_2(\mathscr{V}) = 4$ in the case of a star network, and every edge of the network is in the same relations $\mathscr{R}_{1}$ with the node $\nu_5$ and $\mathscr{R}_{0}$ with the rest of the nodes.


As the second example, consider a cycle (Fig. \ref{fig:cycle}), whose edge set is given by $\mathscr{E}$ $=$ $\{ \epsilon_i, i \in \mathbb{N}_5 \}$, where for $i \in \mathbb{N}_4$, $\epsilon_i = ( \nu_i , \nu_{i+1} )$ and $\epsilon_5 = (\nu_1,\nu_5)$. Associate with every vertex $\nu_i$ a row vector $\mathbf{d}_i$ with five columns, whose element $[\mathbf{d}_i]_{j}, j \in \mathbb{N}_5$ is equal to $k$  if $(\nu_i,{\epsilon}_j) \in \mathscr{R}_{k}$ for some $k \in \mathbb{N}_4\cup\{0\}$. Let $D\in\mathbb{R}^{5 \times 5}$ be the matrix whose $i-$th row is equal to $\mathbf{d}_i$ for all $i \in \mathbb{N}_5$. Then $D$ is given by:

\begin{equation}
D = \left(\begin{array}{rrrrr} 0 & 4 & 3 & 2 & 1 \\  1 & 0 & 4 & 3 & 2  \\ 2 & 1 & 0 & 4 & 3  \\  3 & 2 & 1 & 0 & 4  \\ 4 & 3 & 2 & 1 & 0 \end{array}\right).
\label{eq:G1_laplacian}
\end{equation}

\begin{figure}[ht]
\centering
\includegraphics[width = 0.48\textwidth]{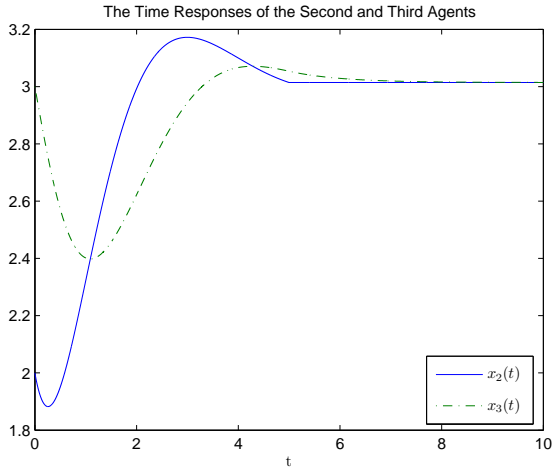}
\caption{The output responses $x_2(t)$ and $x_3(t)$ are plotted for $0\leqslant t \leqslant 10$. The link failure happens at $t_f = 5$, where there is a break in the plot of $x_2(t)$ but not of $x_3(t)$.}
\label{fig:responses}
\end{figure}

\begin{figure}[ht]
\centering
\includegraphics[width = 0.48\textwidth]{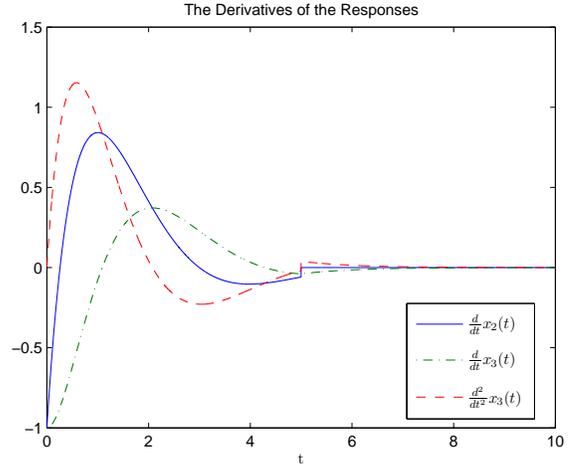}
\caption{The derivatives of $x_2(t)$ and $x_3(t)$ are plotted for $0 \leqslant t \leqslant 10$. At the time of failure $t_f = 5$, there are jump discontinuities in the plots of $\frac{\textrm{{d}}}{\textrm{{d}}t}x_2(t)$ and $\frac{{\textrm{{d}}}^{2}}{\textrm{{d}}t^{2}}x_3(t)$. The plot of $\frac{{\textrm{{d}}}^{2}}{\textrm{{d}}t^{2}}x_2(t)$ contains  an impulse at $t_f = 5$, while $\frac{\textrm{{d}}}{\textrm{{d}}t}x_3(t)$ is a continuous function of time.} 
\label{fig:derivatives}
\end{figure}

It is evident from matrix $D$ that any two distinct vertices will offer a solution $\mathscr{M}_I$ to Problem \ref{prob:detection}, since the locations of $0$ entries do not overlap for distinct vertices. Thence, the designer can detect the failure of any links in the cycle network by observing the jumps in the first four derivatives of any two nodes in the network. In the same vein, any set of two distinct vertices can also be used to uniquely determine which link has failed based on the observed jumps in the first four derivatives. For instance, taking $\mathscr{M}_2 = \{\nu_2,\nu_3\}$, $\epsilon_1$ is the only edge whose failure will produce a jump in the first and second derivatives of $x_2(t)$ and $x_3(t)$, respectively, at the time of failure,  $t = t_f$. Figs \ref{fig:responses} and \ref{fig:derivatives} depict the responses of the second and third agents as well as their derivatives, for a directed cycle of length five initialized at $\mathbf{x}(0) = (1, 2, 3,  4, 5)^{T}$, where for all $t\in \mathbb{R}$, $\mathbf{x}(t) = (x_1(t), x_2(t), x_3(t),  x_4(t), x_5(t))^{T}$. The edge $\epsilon_1$ is removed at time $t_f  = 5$, and as a result, the first derivative of $x_2(t)$ and the second derivative of  $x_3(t)$ exhibit jump discontinuities at $t_f  = 5$.

The above generalizes for finite cycles and stars of arbitrary size. In particular, there is no solution to the isolation problem for any star network and detection can be achieved by observing the first derivative of the common head vertex. For a cycle on the other hand, if all derivatives upto one less than the the network size are observed, then any two nodes offer a solution to not only the detection, but also the isolation problem.

\section{Conclusions}\label{sec:conc}

In this paper, a method was developed, both analytically and algorithmically, that enables the designer of a multi-agent system to detect and isolate single link failures, based on the observed jumps in the derivatives of the output responses of a subset of nodes under the Laplacian consensus dynamics. Two theorems were presented, which relate the jumps in the derivatives at the time of failure to the distance of the the failed link from the observation point. In the proofs, the graph Laplacian can be replaced with any well-defined in-weighting on the graph, so that the extension to general (non-Laplacian) linear dynamics is possible. Any sufficiently regular nonlinear dynamics can be linearized at point $t = 0$ (the time of link failure), thence a generalization to nonlinear network dynamics is also foreseeable. In the future, the authors hope to formalize such extensions.

\bibliographystyle{IEEEtran}
\bibliography{refDistinguishability}

\begin{thebibliography}{1}
\providecommand{\url}[1]{#1}
\csname url@samestyle\endcsname
\providecommand{\newblock}{\relax}
\providecommand{\bibinfo}[2]{#2}
\providecommand{\BIBentrySTDinterwordspacing}{\spaceskip=0pt\relax}
\providecommand{\BIBentryALTinterwordstretchfactor}{4}
\providecommand{\BIBentryALTinterwordspacing}{\spaceskip=\fontdimen2\font plus
\BIBentryALTinterwordstretchfactor\fontdimen3\font minus
  \fontdimen4\font\relax}
\providecommand{\BIBforeignlanguage}[2]{{%
\expandafter\ifx\csname l@#1\endcsname\relax
\typeout{** WARNING: IEEEtran.bst: No hyphenation pattern has been}%
\typeout{** loaded for the language `#1'. Using the pattern for}%
\typeout{** the default language instead.}%
\else
\language=\csname l@#1\endcsname
\fi
#2}}
\providecommand{\BIBdecl}{\relax}
\BIBdecl

\bibitem{mesbahiBook}
M.~Mesbahi and M.~Egerstedt, \emph{{Graph Theoretic Methods in Multiagent
  Networks}}.\hskip 1em plus 0.5em minus 0.4em\relax Princeton University
  Press, 2010.

\bibitem{aminAutomatica}
M.~A. Rahimian and A.~G. Aghdam, ``Structural controllability of multi-agent
  networks: Robustness against simultaneous failures,'' \emph{Automatica},
  2013, in press.

\bibitem{Kleinberg}
J.~Kleinberg, M.~Sandler, and A.~Slivkins, ``Network failure detection and
  graph connectivity,'' \emph{SIAM Journal on Computing}, vol.~38, no.~4, pp.
  1330--1346, 2008.

\bibitem{RenBeard2008}
W.~Ren and R.~Beard, \emph{{Distributed Consensus in Multi-vehicle Cooperative
  Control}}.\hskip 1em plus 0.5em minus 0.4em\relax Springer, 2008.

\bibitem{FailureDetectionACC2012}
M.~A. Rahimian, A.~Ajorlou, and A.~G. Aghdam, ``Characterization of link
  failures in multi-agent systems under the agreement protocol,'' in
  \emph{Proceedings of the American Control Conference}, 2012, pp. 5258--5263.

\bibitem{FailureDetectionCDC2012}
------, ``Detectability of multiple link failures in multi-agent systems under
  the agreement protocol,'' in \emph{Proceedings of the 51st IEEE Conference on
  Decision and Control}, 2012, pp. 118--123.

\bibitem{BiggsGraphTheory}
N.~Biggs, \emph{{Algebraic Graph Theory}}.\hskip 1em plus 0.5em minus
  0.4em\relax Cambridge University Press, 1994.

\bibitem{Dhillon03sensorplacement}
S.~S. Dhillon and K.~Chakrabarty, ``Sensor placement for effective coverage and
  surveillance in distributed sensor networks,'' in \emph{Proceedings of IEEE
  Wireless Communications and Networking Conference}, 2003, pp. 1609--1614.

\bibitem{Leskovec}
J.~Leskovec, A.~Krause, C.~Guestrin, C.~Faloutsos, J.~VanBriesen, and
  N.~Glance, ``Cost-effective outbreak detection in networks,'' in
  \emph{Proceedings of the 13th ACM SIGKDD international conference on
  Knowledge discovery and data mining}, ser. KDD '07.\hskip 1em plus 0.5em
  minus 0.4em\relax New York, NY, USA: ACM, 2007, pp. 420--429.

\end{thebibliography}

\end{document}